# Phase stability and the effect of lattice distortions on electronic properties and half-metallic ferromagnetism of Co$_2$FeAl Heusler alloy: An ab initio study

Aquil Ahmad[a)], S. K. Srivastava, and A. K. Das[*]

*Department of Physics, Indian Institute of Technology Kharagpur, India-721302*

E-mails: [*]amal@phy.iitkgp.ernet.in; [a)]aquil@phy.iitkgp.ac.in

**Abstract**

Density functional theory calculations within the generalized gradient approximation (GGA) are employed to study the ground state of Co$_2$FeAl. Various magnetic configurations are considered to find out its most stable phase. The ferromagnetic ground state of the Co$_2$FeAl is energetically observed with an optimized lattice constant of 5.70 Å. After that, the system was subjected under uniform and non-uniform strains, to see their effects on spin polarization (P) and half-metallicity. The effect of spin-orbit coupling is considered in the present study. Half-metallicity (and 100 % P) is retained only under uniform strains started from 0 to +4%, and dropped rapidly from 90% to 16% for the negative strains started from -1% to -6%. We find that the present system is much sensitive under tetragonal distortions as half-metallicity (and 100% P) is preserved only for the cubic case. The main reason for the loss of half-metallicity is due to the shift of the bands with respect to the Fermi level (E$_F$). We also discuss the influence of these results on spintronics devices.

## 1. Introduction

The first half-metallic ferromagnet, based on the Heusler family, was proposed by de Groot et al., in 1983 [1]. Since then, this family created a huge interest in the scientific community due to its potential in spintronic devices [2, 3]. Heusler compounds such as Co$_2$MnX (X= Ge, Si, Sn) [4], Co$_2$MnZ (Z= main group elements) [5], Co$_2$MnZ (Z=Si, Ge) [6], Fe$_2$CoAl [7, 8] and Fe$_2$YAl (Y = Ni, Mn, Cr) [9] attracted enormous interest of the researchers. The interesting feature of these materials is that they exhibit metallic nature for one spin channel and semiconducting nature in other spin channels; hence attribute 100% spin polarization at the Fermi level ($E_F$). This feature can be exploited in (1) spin injection devices [10] with large magnetoresistance (MR) and (2) perfect spin-filters [11]. Additionally, these materials have shown their potentiality for magnetocaloric application [12-14]. Initially, researchers investigated various physical properties of Heusler alloys (HAs) such as non-local spin ordering [15], magneto-optical properties [16]. Later, their focus was on the origin of the half-metallic energy gap [17-19], and spin-orbit (SO) interaction [20]. There are some compounds, which are non-Heusler (see refs. [21, 22]), exhibit half-metallicity. However, Heusler compounds are still required attention due to their novel properties such as high magnetic moment, high Curie temperature (up to 1000 °C), and low coercivity (for ref., see our previous exp. work on $Co_2FeAl$ alloy [23], and references therein). A 100% P may be achieved under some careful conditions, as few results expect a symmetry break in highly ordered surfaces, e.g., NiMnSb/CdS interfaces [24]. Extensive efforts have been devoted for achieving a direct measurement of spin polarization by means of spin-polarized tunneling [25], Andreev reflection technique [26], and spin-polarized photoemission [27]. Unfortunately, the reported value was below 100%. In contrast, half-metallicity was supported by some experiments viz. infrared reflectance spectroscopy [28], and spin-resolved positron annihilation experiment [29]. The high value of magnetoresistance (MR) was not observed in spin-valve using Heusler layers [30]. However, it was observed in powder compact form [31], suggesting that the high value of spin polarization in thin films is difficult to achieve. The electronic structure calculations suggest structural defects [32] and atomic site disorder [33], which reduces the half-metallic (HM) character of Heusler alloys. The effect of structural distortions on the electronic and magnetic properties of some Heusler alloys has been studied in detail [34-36]. Though, the effect of structural deviation (or distortion) from an ideal (i.e., cubic) structure upon half-metallicity and spin polarization (P) is not understood sufficiently, specifically in $Co_2FeAl$ (CFA).

In this paper, first time, we do a systematic study of ground-state properties of CFA employing phase stability under various magnetic states viz. paramagnetic (PM) or non-magnetic (NM), ferromagnetic (FM), ferrimagnetic (FiM) and antiferromagnetic (AFM), to find out the most stable ground state. After that, the system was exposed under uniform and non-uniform strains by means of the lattice constant related to zero (or unstrained) pressure, to see its effects on P and half-metallicity. The present research is highly instructive in synthesizing CFA based thin films, nanostructures and/or heterostructures. Because of only a few percent deviations from the bulk lattice constant results in a loss of 100% P. The present study further proves that non-uniform strains are not the only reason for low-performance of Heusler alloys in the spin-based devices.

**2. Computational details:**

Band structure calculations have been performed using the Wien2k computational code [37] based on density functional theory (DFT). The accuracy of the electronic structures calculation results strongly depends upon the choices of exchange-correlation functional. Previous studies suggest that general gradient approximation (GGA) is more appropriate for the strongly correlated d-f electron systems viz. half-metals [38-40]. Consequently, we use generalized gradient approximation (GGA) [41] of Perdew-Burke-Ernzerhof (PBE) in our calculations. The electronic configurations used for the valence states of Co, Fe, and Al are: $3p^6, 4s^2, 3d^7$; $3p^6, 4s^2, 3d^6$, and $3s^2, 3p^1$ respectively. The non-spherical contribution of the charge density was being considered up to $l_{max} = 10$ within the muffin-tin (MT) sphere. The muffin-tin sphere radii were chosen as 2.28 a.u. for Co/Fe and 2.15 a.u. for Al atoms resulting in nearly touching spheres. The cut-off parameter $R_{MT} \times K_{max} = 7$ was set for all calculations. The charge density and potential may expand in the interstitial region up to $G_{max}=12$ (a.u$^{-1}$). The grid of 15×15×15 mesh was used during calculations. The spin-orbit coupling (SOC) effect is considered here. The total energy versus volume curve was fitted with the Birch-Murnaghan equation of state [42], to give the optimized parameters.

**3. Results and discussion**

**3.1. Phase stability**

Theoretical investigations have been performed to study the physical properties of the CFA. Our primary objective was to find out the most stable ground state and study its limitation to half-metallicity, spin polarization, Slater-Pauling (SP) rule and magnetic properties, when the system undergoes some structural distortions (i.e., deviation from its ideal structure of Cu$_2$MnAl prototype). Half-metallic ferromagnets (HMFs) often exhibit 100 % P, at the E$_F$, are crucial for their applicability and performances in spintronic devices. It is known that Co$_2$FeAl can be crystallized in regular L2$_1$ structure (see figure 1(a)) under Fm$\bar{3}$m space group (# 225), where all the atoms belong to four Wyckoff sites: 4a (0, 0, 0), 4c (0.25, 0.25, 0.25), 4b (0.5, 0.5, 0.5), and 4d (0.75, 0.75, 0.75). In FCC lattice, antiferromagnetic (AFM) ordering is difficult to achieve owing to the presence of geometrical frustration in the system; hence, structural distortion is essential. Tetragonal distortion usually occurs in Heusler alloys [43]. Therefore, to obtain an AFM structure, we construct a supercell of CFA, which takes the space group Pmmm (# 47), as shown in figure 1. Here, the ferromagnetic (FM) planes of Co spins are alternatively arranged in a specific direction [001]. Similarly, FM-planes of Fe spins are also arranged (see Fig. 1(b, c)) [45].

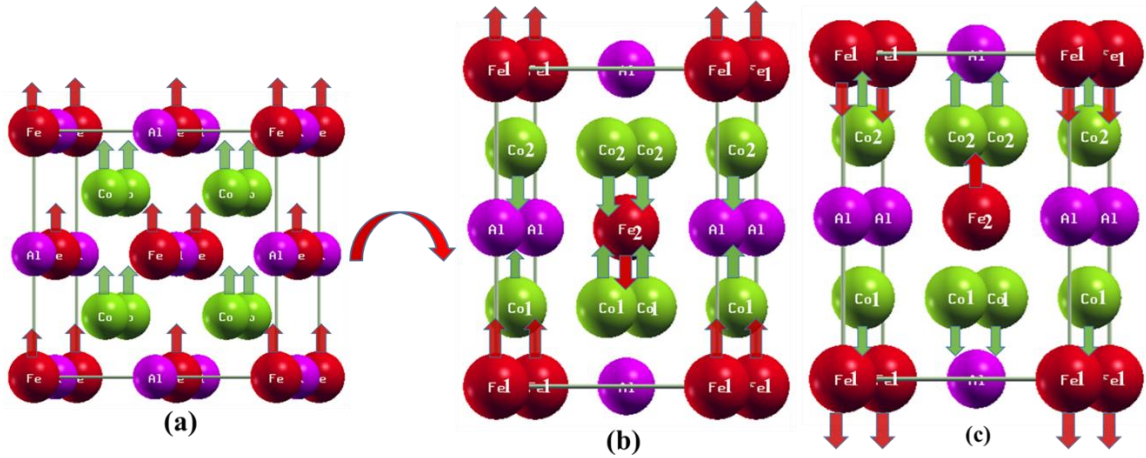

**Figure 1**. (Color online) Crystal structures of $Co_2FeAl$ alloy under (a) FM, (b) AFM-I and (c) AFM-II ordering. All crystal structures have been generated using XCrysDen software [44].

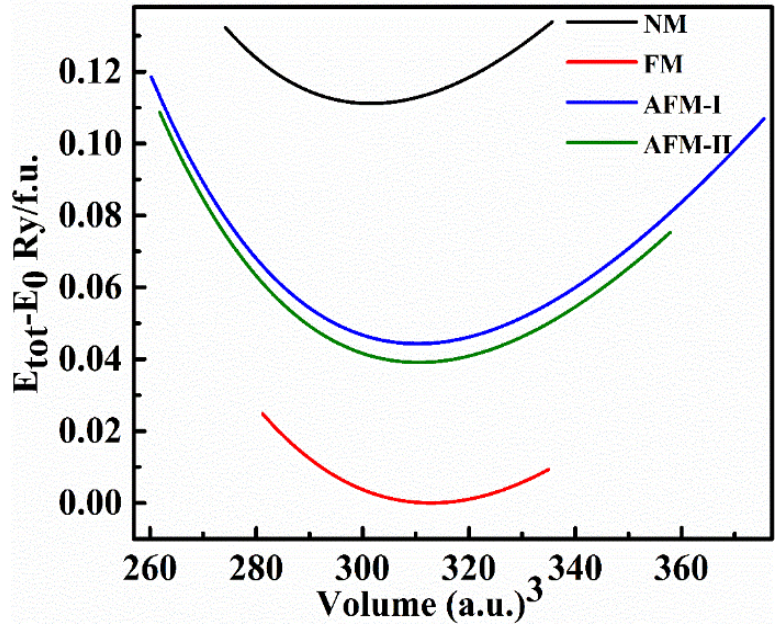

**Figure 2.** (Color online) Total energy difference ($E_{tot}$-$E_0$) as a function of the unit cell volume of $Co_2FeAl$ under NM or PM, FM, and AFM configurations. The optimized curves are obtained after fitting with the Birch-Murnaghan equation of state.

To obtain the ground state of CFA, lattice optimizations have been performed in different magnetic states viz. paramagnetic (PM), ferromagnetic (FM), ferrimagnetic (FiM) and antiferromagnetic (AFM). The total energy difference versus volume [($E_{tot}$-$E_o$)-V)] curves are presented in figure 2. Note that the ferrimagnetic state was not achieved even after more than 100 iterations in the self-consistent field (scf) cycles.

| Parameter | NM (Cubic) | FM (Cubic) | AFM-I (Tetragonal) | AFM-II (Tetragonal) |
|---|---|---|---|---|
| lattice constants (Å) | 5.72 (previous exp.)* [23], 5.73 (theory)** [46], | | | |
| Eq. lattice constants (Å) | 5.63 | 5.70, 5.69 [43] | a = b = 4.02 c = 5.69 | a = b = 4.02 c = 5.69 |
| Bulk modulus B (GPa) | 211.69 | 190.64 (this work) (a) 190.19 [calcul.] (b) 239.0 [calcul.] (c) 204.0 [exp.] | 193.35 | 193.50 |
| Derivative of Bulk modulus (B´) | 4.45 | 4.60 (a) 4.55 | 4.74 | 4.73 |
| Total energy ($E_0$) | -8605.1407 | -8605.2518 | -8605.2075 | -8605.2127 |

* At room temperature (RT); (a) Ref. [38, 47]; (b) Ref. [48]; (c) Ref. [49]

** At T = 0 K

**Table 1.** The calculated optimized lattice parameter $a_0$ (Å), bulk modulus B (GPa) and its pressure derivative B´, equilibrium volume $V_0$, and total energy ($E_0$) per cell of the $Co_2FeAl$. The other results are presented in parenthesis.

The Birch-Murnaghan fitted parameters such as optimized lattice constant ($a_0$), bulk modulus ($B_0$) along with its pressure derivate (B´), and the total energies per cell ($E_0$) are gathered in table 1 in conjunction with those of the other results shown in brackets for comparison. From figure 2, we established the stability of the ferromagnetic state where we have imposed the CFA system to be either ferrimagnetic, antiferromagnetic, or paramagnetic. On a comparison of the total energy, it is clear that $Co_2FeAl$ energetically prefers FM ordering as its ground state, and hence likely to be observed in the experiments. The optimized lattice constant ($a_{opt.}$) of FM type CFA was found to be 5.70 Å, which closely matches with the experimental one (see table1). This observation is consistent with the experimental results, where the FM ground state of CFA, has earlier been reported [23, 50]. To date, there are no experimental reports on AFM or FiM ground state of CFA. Figure 3 shows the spin-polarized total and atomically resolved density of states (DOS). It is clear that the density of states near the Fermi level ($E_F$), are dominated by 3d states of Co and Fe atoms. Whenever the majority states are nearly fully occupied, the two peaks in minority states, just above the $E_F$ are due to Co and Fe 3d contributions. The broad structure in the lowest energy region between -8.0 and -5.8 eV (not shown) is due to Al (non-magnetic) 3s, and 3p states, which are very well separated from 3 d states of Co/Fe found between -5.3 to 4 eV. The Fermi-level is falling in the gap for minority states, attributing 100% P, at the $E_F$. It is well described in ref. [17] that for such half-metallic compounds, the total magnetic moment should be an integer.

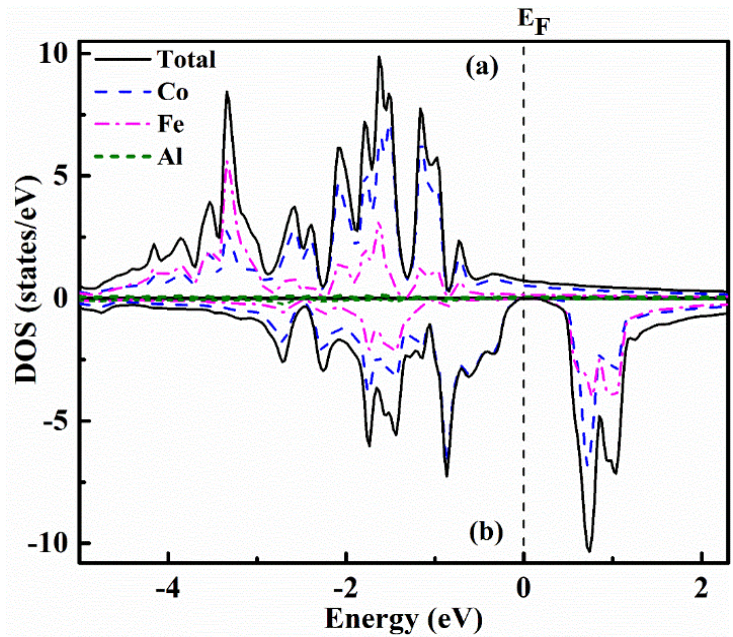

**Figure 3.** The total and partial DOS of $Co_2FeAl$ are shown of (a) majority (up), and (b) minority (down) spins. The Fermi level ($E_F$) was set at zero energy.

| Structure | $M_{Co}$ ($\mu_B$) | $M_{Fe}$ ($\mu_B$) | $M_{Al}$ ($\mu_B$) | $M_{Tot}^{Cal}$ ($\mu_B$) | $M_{Tot}^{Rep}$ ($\mu_B$) | $M_{Tot}^{Exp}$ ($\mu_B$) | $M_{Sp}$ ($\mu_B$) |
|---|---|---|---|---|---|---|---|
| FM | 1.23 | 2.81 | -0.05 | 5.00 | 5.08 [43] 4.99 [51] | 5.2 (bulk) [52] 5.3-6.5 [23] (previous exp.) | 5.0 [17] |
| AFM-I | Co1/Co2 -0.13/+0.13 | Fe1/Fe -2.91/+2.91 | Al1/Al2 -0.01/+0.01 | 0.0 | | | |
| AFM-II | Co1/Co2 -0.12/ +0.12 | Fe1/Fe2 +2.91/-2.91 | +0.01/-0.01 | 0.0 | | | |

**Table 2.** The calculated total and partial magnetic moments in FM, AFM-I, and AFM-II states: ($M_t$), $M_{Co}$, $M_{Fe}$, and $M_{Al}$ are listed. The other theoretical and experimental results for comparison are shown with references in brackets.

From the self-consistent field (scf) calculation results (see table. 2), the total magnetic moment per cell of FM ordered CFA alloy is found to be 5.0 $\mu_B$, which is consistent with the Slater-Pauling (SP) rule [17] and hence, resulting in a perfect half-metal. It is clear that only Co and Fe atoms contributed in the total magnetic moment, and Al has a negligible moment. Thus the FM interaction between Co-Fe is the strongest bonding interaction, determining the energy gap of 0.11 eV in the minority-spin band. Due to the covalent hybridization between Co and Fe, bonding and antibonding states are formed, which also determine the position of the Fermi level ($E_F$) [31].

The Fe atom has the largest magnetic moment, and it coupled ferromagnetically with the Co atom. Spin down states essentially represent the characteristic of Co and Fe atoms; therefore, it is realistic to consider the hybridization among them. The four sp bands are situated far below the $E_F$ hence, inappropriate for the gap. Consequently, the hybridization of the 15 d-states of the two Co atoms and one Fe atom is being considered. The hybridization between d-orbitals of the two Co and one Fe atoms, assuming that the coordination of the Co atoms is octahedral, is schematically shown in figure 4, to explain the reason for the band gap in $Co_2FeAl$. First, we sketch the hybridization between the Co atoms, as shown in figure 4 (left side). The five d-orbitals of Co comprise of the 3-fold degenerate *dxy*, *dyz*, and *dxz*, and the 2-fold degenerate $dz^2$, and $dx^2-y^2$ states. The $t_{2g}$ ($e_g$) orbitals of Co atoms can only pair with the $t_{2g}$ ($e_g$) orbitals of the other Co atom. The $t_{2g}$ and $e_g$ are bonding orbitals, while $t_{1u}$ and $e_u$ are antibonding orbitals. The degeneracy of the orbitals is represented by the numbers providing the respective orbitals. Now the hybridization scheme is presented between the Co-Co hybridized orbitals and the Fe d-orbitals, as shown in figure 4 (right side). The doubly degenerate $e_g$ orbitals hybridize with the $d_4$ (or $d_z^2$) and $d_5$ ($dx^2-y^2$) orbitals of Fe and form doubly degenerate bonding and antibonding $e_g$ orbitals.

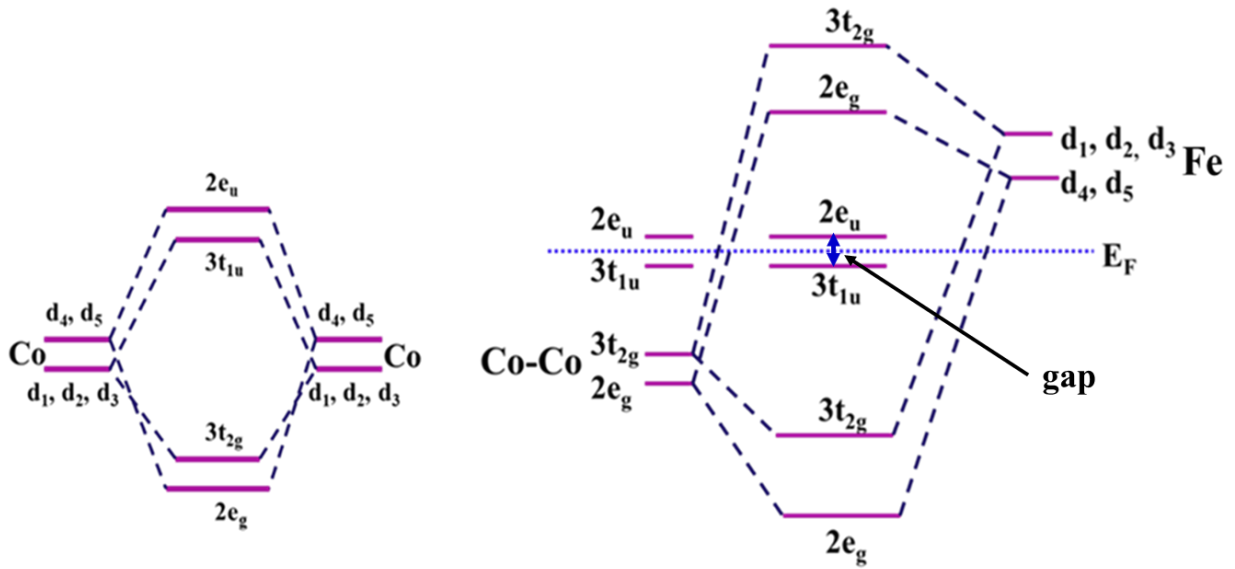

**Figure 4.** A schematic diagram of possible hybridizations between d orbitals located at different sites in the case of the $Co_2FeAl$ compound. For simplicity, *dxy*, *dyz* and *dxz* orbitals are represented by $d_1$, $d_2$ and $d_3$, and $dz^2$ and $dx^2-y^2$ orbitals are represented by $d_4$ and $d_5$, respectively. The degeneracy of the corresponding orbital is represented by the coefficients. The $t_{2g}$ and $e_g$ states represent the bonding, and $t_{1u}$, $e_u$ represent the antibonding orbitals.

The triply degenerated $t_{2g}$ orbitals couple with the $d_1$, $d_2$, and $d_3$ orbitals of Fe and form six new hybrid orbitals, three of which are bonding and the other 3 are antibonding. Lastly, the doubly and triply degenerate $e_u$ and $t_{1u}$ orbitals can't couple with any d-state of Fe, as these states are orthogonal to $e_u$ and $t_{1u}$ states of Co. The $t_{1u}$ states lie below the Fermi energy, whereas $e_u$ states lie above $E_F$. Therefore, out of 15, eight states are filled, and the rest are empty. The $E_F$ falls between the five non-bonding Co states in such a way that the three $t_{1u}$ bands are fully occupied, and the rest 2-bands of $e_u$ are empty. The interaction between Co-Co atoms actually determines the real gap in $Co_2FeAl$, forming due to the presence of the splitting of $e_u$ and $t_{1u}$ states near the $E_F$ [53].

### 3.2. Effect of lattice distortions on the electronic and magnetic properties

In this section, we study the effect of lattice distortions on half-metallic ferromagnetism of $Co_2FeAl$ alloy. Initially, the calculation was started from zero strain, i.e., at equilibrium lattice constant, $a_{(optim.)} = 5.70$ Å. Then strains applied to the system are -6%, -5%, -4%, -3%, -2%,

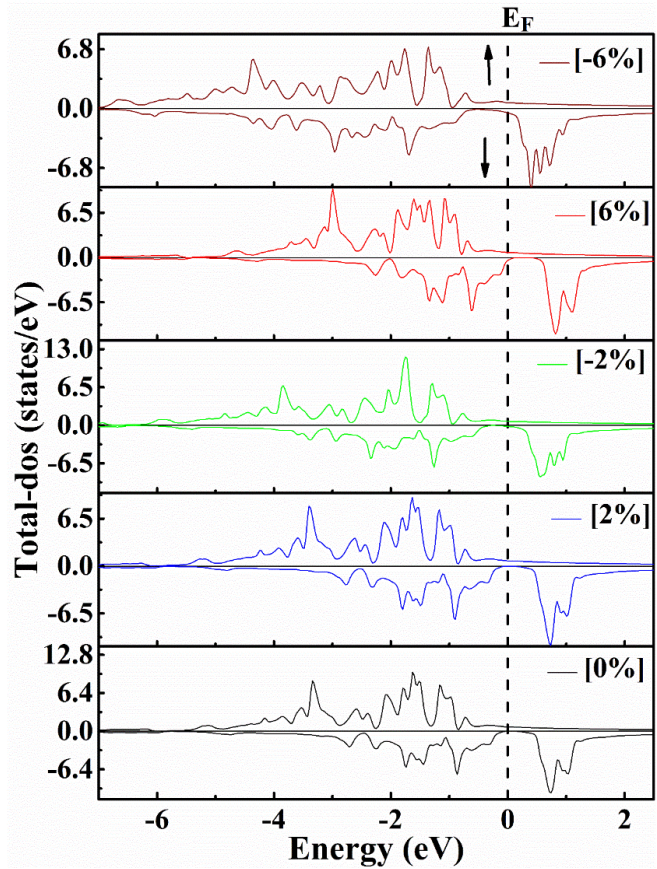

**Figure 5.** The spin-polarized density of states (DOS) plots of $Co_2FeAl$ alloy under uniform strains. The majority and minority spins are shown by up and down arrows. Fermi level (i.e., $E_F = 0$) is shown by a vertical dashed line. Note that the data are shown only for -6%, -2%, 0%, 2%, and 6%.

-1%, up to +6% relative to $a_{(optim.)}$. Generally, in a layered structure, a significant epitaxial strain is expected from its adjacent layers, resulting in non-uniform strain. That is why we also studied the effects of tetragonal distortions with the c axis, which was varied from -6% to +6%, keeping the total volume of the cell constant. We have carefully chosen the ranges of distortions to see the trends of the density of states and their effect on half-metallicity, as these properties are highly sensitive to the distortions applied. Numerous reports on band structures are available on unstrained (i.e., cubic) cases based on the $Co_2FeAl$ alloy [43, 51, 54-60]. However, some of our calculation results must be addressed in predicting the device compatibility under uniform-strain and tetragonal distortions. Additionally, a systematic comparative study on the phase stability under various magnetic and non-magnetic states is performed. Figure 5 shows the effect of the uniform strains on the DOS. The basic structure of the density of states remains the same, which is not astonishing as the basic crystal symmetry is unchanged. Here the main dissimilarity between the DOS's is that the energy gaps shift with respect to the Fermi energy. The gap centers are also slightly changed because of the strains. As we know that for a free electron gas, $E_F \propto V^{-2/3}$, hence a shift in $E_F$ is expected with the unit cell volume. A comparison of all values as obtained from SCF calculations are gathered in table 3. We have calculated spin polarization (P) as the ratio $[D\uparrow(E_F)-D\downarrow(E_F)] / [D\uparrow(E_F)+D\downarrow(E_F)]$, where $D\uparrow(E_F)$ and $D\downarrow(E_F)$ are the majority (i.e., spin-up) and minority (i.e., spin-dn) density of states at the $E_F$. The trends of the total magnetic moment per cell and partial moments of Fe and Co, spin polarization, and the center position and the gap width are shown in figure 6 (a, b, c), respectively. It can be seen that 100% SP is retained only from 0 to +4% (positive) strain and dropped rapidly from 90% to 16% for negative strains started from -1% to -6%. In figure 6, we have also plotted gap centers with respect to the Fermi energy ($E_F$) represented by a point and the energy gap ($E_{gap}$) by (error) bars.

**Table. 3.** SCF+ SO results showing spin polarization, the total magnetic moment per cell, partial magnetic moments of Co and Fe, and half-metallic ferromagnetic (HMF) behavior in case of uniform strain.

| Uniform strain % | $\rho\uparrow$ ($E_F$) | $\rho\downarrow$ ($E_F$) | P % | $M_{Co}$ | $M_{Fe}$ | $M_t$ ($\mu_B$/f.u.) | HMF |
|---|---|---|---|---|---|---|---|
| -6% | 0.67 | 0.48 | 16.52 | 1.167 | 2.577 | 4.76 | No |
| -5% | 0.70 | 0.40 | 27.27 | 1.170 | 2.626 | 4.83 | No |
| -4% | 0.69 | 0.31 | 38 | 1.199 | 2.664 | 4.89 | No |
| -3% | 0.69 | 0.23 | 50 | 1.207 | 2.695 | 4.92 | No |
| -2% | 0.69 | 0.29 | 40.81 | 1.220 | 2.736 | 4.97 | No |
| -1% | 0.73 | 0.05 | 87.17 | 1.215 | 2.773 | 4.96 | No |
| 0% | 0.71 | 0 | 100 | 1.230 | 2.806 | 5.0 | Yes |
| +1% | 0.72 | 0 | 100 | 1.228 | 2.815 | 5.01 | Yes |
| +2% | 0.72 | 0 | 100 | 1.235 | 2.829 | 5.01 | Yes |
| +3% | 0.73 | 0 | 100 | 1.237 | 2.846 | 5.01 | Yes |
| +4% | 0.74 | 0 | 100 | 1.239 | 2.865 | 5.01 | Yes |
| +5% | 0.75 | 0.10 | 76.47 | 1.248 | 2.882 | 5.05 | No |
| +6% | 0.75 | 0.42 | 28.20 | 1.264 | 2.903 | 5.10 | No |

The energy gap is increased from 0.1 eV to 0.24 eV for the strain between 0 to +6%, and decreased from 0.1 eV to 0.03 eV for the strain between 0 to -6%.

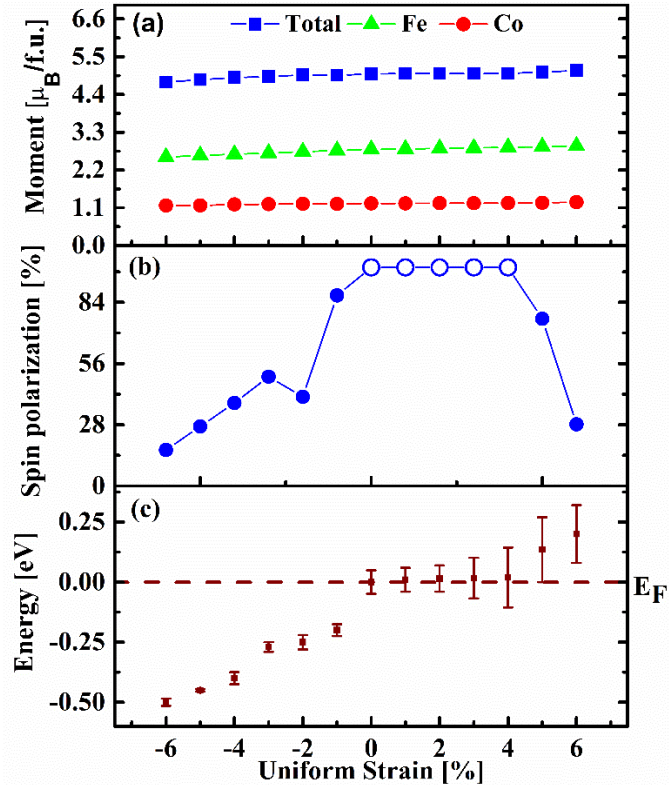

**Figure 6.** Uniform strain effect on the magnetic moment, spin polarization, and band gap for Co$_2$FeAl are shown in (a), (b), and (c), respectively. 100% spin polarization is presented by a hollow circle, the Fermi energy is represented by a horizontal dashed line, and the band gap center is shown by a point. The energy gap (E$_{gap}$) is shown by (error) bars.

The main reason for the loss of the half-metallicity is due to the shift of the bands with respect to the Fermi energy [35]. When negative strain is applied to the lattice, causes it to squeeze. The resulting smaller volume of the unit cell results in broader bands. The dispersions in the electronic states are larger because of the steeper bands. For the positive strain, it is just the opposite. Notice that the above results are with the spin-orbit coupling (SOC). We also noticed no significant difference in total magnetic moments of the cell with or without SOC during the self-consistent field (SCF) calculations. The gap center and width changes can be explained as follows: exchange coupling will be increased when electrons are more localized, and hence, minority bands are shifted at higher with respect to their majority bands. In Figure 6, we can see the variation of the total moments as well as partial moments of the atoms with respect to the strains. From these results, we accomplished that the integer magnetic moment will be the necessary condition to obtain a 100% spin-polarized Heusler alloy.

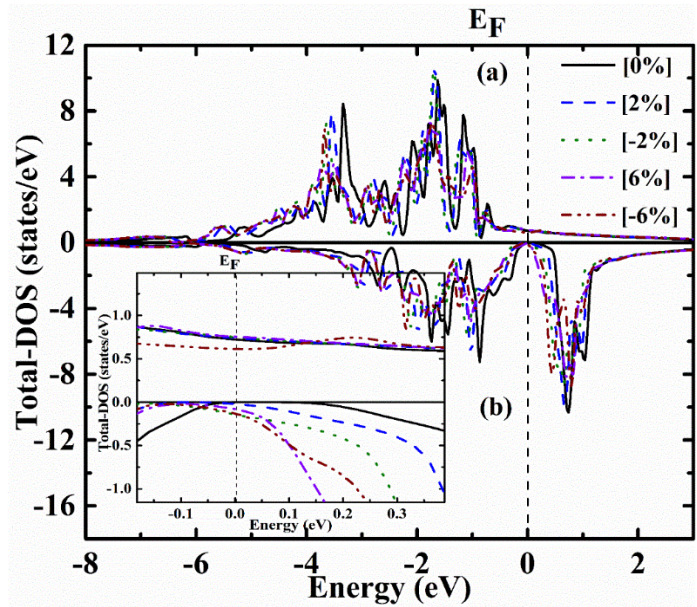

**Figure 7.** The density of states (DOS) plot of Co$_2$FeAl alloy under tetragonal distortions are shown of (a) majority and (b) minority spins. Data are presented for -6%, -2%, 0%, 2%, and 6%. Inset represents the magnified version near the E$_F$.

An unstrained Co$_2$FeAl Heusler alloy has a total magnetic moment of 5.0 µ$_B$, which is quite consistent with the value predicted by the Slater-Pauling (SP) rule. An integer magnetic moment is retained and hence half-metallicity, too, up to when the changes in atomic magnetic moments were compensated by each other. Moreover, until the covalent bonding is strong between transition metals, the half-metallicity is preserved, and when it is weakened, resulting in a loss of half-metallic behavior. µ$_{Co}$ and µ$_{Fe}$ were decreased for negative strains (see figure 6), so the total magnetic moment of the cell is reduced. On the contrary, the total magnetic moment slightly increases for the positive strains (see table 3).

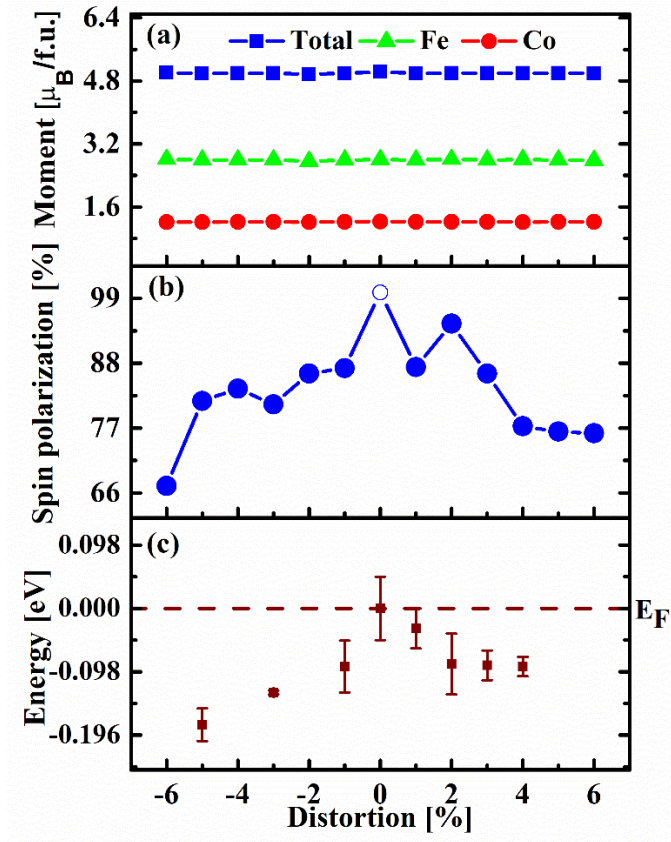

**Figure 8.** Non-uniform strain effect on the magnetic moment, spin polarization, and band gap for Co$_2$FeAl are shown in (a), (b), and (c), respectively. 100% spin polarization is shown by a hollow circle, the Fermi energy is represented by a horizontal line in (c), the band gap center is indicated by a point, and E$_{gap}$ is shown by (error) bars.

The effect of tetragonal distortions upon the total density of states (T-DOS) is presented in figure 7. The general shapes of DOS are not changed. The influence of the tetragonal distortion on $E_{gap}$ and spin-polarization (P) is shown in figure 8. We observed that $Co_2FeAl$ alloy is much sensitive under tetragonal distortions, and 100% spin polarization is only preserved for the unstrained case. The changes in band gaps ($E_{gap}$) are also displaced relative to the applied distortions, which is not monotonic. A comparison of all values is given in table 4. Interestingly, we have noticed that the spin polarization (P) decreases drastically in the case of uniform-strains as compared to the tetragonal distortions. Under uniform-strains, the $E_{gap}$ shifts to the left side of the $E_F$ for the negative strains (see figure 5), and right side of the $E_F$ for the for the positive strains. On contrary, the gaps only shift to the left side of the $E_F$ under tetragonal distortions and gap width decreases, hence, a pseudogap forms causing a loss of half metallic (and 100%P) character in CFA [34]. Tetragonal distortions completely close the gap which might be the reason of non-monotonic behavior of the $E_{gap}$ relative to $E_F$ (see figure 8). The spin polarization is significantly high under tetragonal distortions as compared with the uniform-strain case (see the table 3&4). The reason might be due to the total magnetic moments which were not much deviated from the Slater Pauling rule. Hence we conclude that while half-metallicity (and 100%P) is much sensitive under tetragonal distortions, the spin-polarization values are quite high, which is a good sign for the spintronic devices.

**Table. 4.** SCF+ SO results showing spin polarization, the total magnetic moment per cell, partial magnetic moments of Co and Fe, and half-metallic behavior (HMF) in case of non-uniform strains (tetragonal distortion).

| Distortion % | $\rho\uparrow$ ($E_F$) | $\rho\downarrow$ ($E_F$) | P % | $M_{Co}$ | $M_{Fe}$ | $M_t$ ($\mu_B$/f.u.) | HMF |
|---|---|---|---|---|---|---|---|
| -6% | 0.611 | 0.12 | 67.16 | 1.216 | 2.810 | 5.01 | No |
| -5% | 0.69 | 0.07 | 81.57 | 1.214 | 2.791 | 4.99 | No |
| -4% | 0.79 | 0.07 | 83.70 | 1.217 | 2.794 | 4.99 | No |
| -3% | 0.76 | 0.08 | 81.02 | 1.219 | 2.793 | 4.99 | No |
| -2% | 0.75 | 0.14 | 86.23 | 1.212 | 2.762 | 4.96 | No |
| -1% | 0.73 | 0.05 | 87.17 | 1.220 | 2.791 | 4.99 | No |
| 0% | 0.71 | 0 | 100.0 | 1.230 | 2.806 | 5.0 | Yes |
| +1% | 0.74 | 0.05 | 87.34 | 1.220 | 2.792 | 4.99 | No |
| +2% | 0.74 | 0.02 | 94.71 | 1.222 | 2.808 | 4.99 | No |
| +3% | 0.74 | 0.05 | 86.23 | 1.220 | 2.791 | 4.99 | No |
| +4% | 0.75 | 0.09 | 77.30 | 1.214 | 2.804 | 4.99 | No |
| +5% | 0.75 | 0.10 | 76.47 | 1.218 | 2.790 | 4.99 | No |
| +6% | 0.74 | 0.10 | 76.10 | 1.218 | 2.783 | 4.99 | No |

The loss of 100% P due to uniform-strain and the distortions will affect the device performance, e.g., thin films prepared by sputtered technique may have stresses of ± 1 GPa because of the preparation method alone [61]. Therefore, to get optimal results, vigilant control of deposition techniques are required. Based on our results and discussions, the following would be the criteria of choosing a good spacer layer: (i) it should be of good lattice-match with the Heusler alloys and (ii) a good band matching as well as the long diffusion length would be the essential parameter for obtaining the expected results in spintronic devices.

## 4. Conclusion

In summary, we conclude that $Co_2FeAl$ alloy energetically prefers a ferromagnetic ground state. The integer magnetic moment is a necessary criterion to obtain a half-metal. Our study suggests that a careful choice of substrate or spacer layer is mandatory to achieve expected results in spin-based devices. We have demonstrated that the half-metallicity (HM) of $Co_2FeAl$ is much sensitive to the distortions (uniform/non-uniform) applied. The reason for the loss of half metallicity (i.e., 100% spin polarization) is due to the shift of the bands with respect to the $E_F$. In the case of tetragonal distortions, the closing of the gap is responsible for the loss of half-metallicity due to the weakening of the covalent hybridization between the Co-Fe atoms. We believe that the present study is highly instructive of synthesizing a Heusler based thin films as well as the heterostructures.

**Acknowledgments**

Aquil Ahmad sincerely acknowledges the University Grant Commission (UGC) Delhi, MHRD Delhi, India, for providing fellowship for Ph.D. work. A. K. Das acknowledges the financial support of DST, India (project no. EMR/2014/001026). We also acknowledge our departmental computational facility, IIT Kharagpur, India.

**References**

[1] de Groot R A, Mueller F M, Engen P G and Buschow K H J 1983 Phys. Rev. Lett. **50** 2024-2027

[2] Wolf S, Awschalom D, Buhrman R, Daughton J, Von Molnar S, Roukes M, Chtchelkanova A Y and Treger D 2001 Science **294** 1488-1495

[3] Fang C M, de Wijs G A and de Groot R A, 2002 J. Appl. Phys. **91** 8340

[4] Picozzi S, Continenza A and Freeman A J 2002 Phys. Rev. B, **66** 094421

[5] Ishida S, Fujii S, Kashiwagi S and Asano S 1995 J. Phys. Soc. Japan **64** 2152-2157

[6] Ishida S, Masaki T, Fujii S and Asano S. 1998 Physica B Cond. Matter **245** 1-8

[7] Ahmad A, Mitra S, Biswas S, Srivastava S K and Das A K 2019 AIP Conf. Proc. **2115** 030508.

[8] Ahmad A., Das A. K. and Srivastava S. K. 2020 Eur. Phys. J. B **93** 96.

[9] Dahmane F, Mogulkoc Y, Doumi B, Tadjer A, Khenata R., Bin Omran S., Rai D.P., Murtaza G and Varshney D, J. 2016 Magn. Magn. Mater. **407** 167-174

[10] Datta S. and Das B 1990, Appl. Phys. Lett. **56** 665-667

[11] Kilian K and Victora R 2000 J. Appl. Phys. **87** 7064-7066

[12] Żuberek R, Chumak O, Nabiałek A, Chojnacki M, Radelytskyi I and Szymczak H 2018 J. Alloys Comp. **748** 1-5

[13] Sagotra A K Errandonea D and Cazorla C 2017 Nature Commun. **8** 1-7

[14] Ahmad A, Srivastava S K and Das A K 2019 arXiv:1909.10201

[15] Mohn P and Supanetz E, 1998 Philosophical Magazine B, **78** 629-636

[16] de Groot R A, Mueller F M, Van Engen P.G. and Buschow K.H.J., 1984 J. Appl. Phys., **55** 2151-2154.

[17] Galanakis I, Dederichs P and Papanikolaou N 2002 Phys. Rev. B **66** 174429

[18] Galanakis I, Dederichs P H and Papanikolaou N 2002 Physical Review B, **66** (2002) 134428

[19] Galanakis I 2002 J. Phys. Conden. Matter **14** 6329

[20] Mavropoulos P, Sato K, Zeller R, Dederichs P H, Popescu V and Ebert H 2004 Phys. Rev. B, **69** 054424

[21] Xie W H, Xu Y Q, Liu B G and Pettifor D G 2003 Phys. Rev. Lett., **91** 037204

[22] Gupta A and Sun J 1999 J. Magn. Magn. Mater. **200** 24-43

[23] Ahmad A, Mitra S, Srivastava S K, Das A K 2019 J. Magn. Magn. Mater. **474** 599-604